\documentclass[preprint,aps]{revtex4}
\usepackage{graphicx}
\begin{document}
\title{``Bi-modal'' Isoscalar Giant Dipole Strength in $^{58}$Ni \\}
\author{B.K.~Nayak, U.~Garg, M.~Hedden,
M.~Koss, T.~Li, Y.~Liu, P.V.~Madhusudhana~Rao, S.~Zhu}
\affiliation{
Physics Department, University of Notre Dame, Notre Dame,
IN 46556, USA \
}
\author{M.~Itoh, H.~Sakaguchi, H.~Takeda, M.~Uchida,
Y.~Yasuda, M.~Yosoi}
\affiliation{
Department of Physics, Kyoto University, Kyoto 606-8502,
Japan \
}
\author{H.~Fujimura, M.~Fujiwara, K.~Hara}
\affiliation{
Research Center for Nuclear Physics, Osaka University,  Mihogaoka 10-1 Ibaraki, Osaka 567-0047, Japan \
}
\author{T.~Kawabata}
\affiliation{
Center for Nuclear Study, Graduate School of Science, University of Tokyo, Bunkyo, Tokyo 113-0033, Japan \
}
\author{H.~Akimune}
\affiliation{
Department of Physics, Konan University, Kobe, Hyogo 658-8501, Japan \
}
\author{M.N.~Harakeh}
\affiliation{
Kernfysisch Versneller Instituut, University of Groningen, 9747 AA Groningen, The Netherlands \
}

\vskip 8 mm
\date{\today}
\vspace*{2 cm}

\begin{abstract}

The strength distribution of the isoscalar giant dipole resonance (ISGDR) in
$^{58}$Ni has been obtained over the energy range 10.5--49.5 MeV
via extreme forward angle scattering (including 0$^{\circ}$) of 386 MeV $\alpha$ particles.
We observe a ``bi-modal'' ${E1}$  strength distribution
for the first time in an A $<$ 90 nucleus. The observed ISGDR
strength distribution is in good agreement
with the predictions of a recent RPA
calculation.


\end{abstract}

\maketitle

The compressional-mode giant resonances in the atomic nuclei---the
isoscalar
giant monopole resonance (ISGMR) and the isoscalar giant dipole resonance
(ISGDR)---provide a direct method to obtain
the incompressibility of the nucleus and of nuclear matter
($K_{nm}$) \cite{hara}.  Although ISGMR has been investigated
extensively for a large number of nuclei in the past, the exotic
ISGDR has been identified only in a few nuclei and the location of
ISGDR is not systematically established over the wide mass region.
One major concern with ISGDR data had been that the nuclear
incompressibility extracted from the centroid of the ISGDR strength
distribution was significantly different from that obtained from the
ISGMR data. In recent work, this ambiguity has been resolved
for $^{208}$Pb by a more precise, background-free measurement of
ISGDR strength distribution, and the value of $K_{nm}$ obtained
from the ISGMR data is now consistent with that from the ISGDR data for
$^{208}$Pb \cite{uchida1}.

The experimentally-observed ISGDR strength distribution in all A $\ge$ 90
nuclei has a ``bi-modal'' structure \cite{uchida2,itoh,dhy1,dhy2},
in agreement with
predictions of recent theoretical work \cite{colo,dario,jorge,shlomo}. Of
these, only the high-energy (HE)
component depends on $K_{nm}$ and, hence, is of interest from the point of
view of determining an experimental value for this important parameter.
The low-energy (LE) component, which is quite small in
comparison with the HE component, is located much
higher in excitation energy than the expected $1 \hbar\omega$
component of the ISGDR, previously identified by
Poelhekken {\it et al.}\cite{isgdr1}.
As well, it is lower in energy than the isovector giant dipole resonance
(IVGDR) which can be excited in inelastic $\alpha$ scattering via Coulomb
excitation; in the event, the full
expected IVGDR strength is subtracted out in the analysis of all
aforementioned data.
The exact nature of this component is not fully understood yet, 
although suggestions have been made that it might represent the ``toroidal''
or ``vortex'' modes; Refs. \cite{garg,colo1} provide a review of the
recent experimental and theoretical results on ISGDR. 

For  $^{58}$Ni, there has been only one
recent measurement, wherein a concentrated ISGMR and isoscalar giant
quadrupole resonance (ISGQR) strength distribution has been observed,
but  the ISGDR strength is reported to be spread more or less
uniformly over $\it{E}_{\rm{x}}$ = 12 to 35 MeV \cite{lui}. This
observation leaves a few open questions: Is the ISGDR
strength fragmented in light nuclei such as $^{58}$Ni? Are we
missing the resonance strength distribution because of experimental
limitations? In an attempt to answer these questions, we have carried out
measurements on excitation of isoscalar giant resonances in $^{58}$Ni.
In this Letter, we report our results on the ISGDR strength distribution
in $^{58}$Ni. We find that the ISGDR in this nucleus has a ``bi-modal''
structure as well, similar to that in the medium- and heavy-mass nuclei,
and that the experimentally observed ISGDR strength is in good
agreement with  predictions of a recent RPA calculation.

  The $^{58}$Ni$(\alpha, \alpha^\prime)$ experiment at $E_{\alpha}$ = 386
MeV was performed at the ring-cyclotron facility of Research Center
for Nuclear Physics (RCNP), Osaka University. Details of the experimental
measurements and data analysis procedures have been provided in Refs.
\cite{itoh,uchida1};
only the salient points are elaborated upon below. $\alpha$-particles,
inelastically scattered off a 5.8 mg/cm$^{2}$-thick $^{58}$Ni target,
were momentum analyzed in the spectrometer, Grand Raiden \cite{fuji}, and
detected in the focal-plane detector system comprised of two
multi-wire drift-chambers and two scintillators, providing particle
identification as well as the trajectories of the scattered particles. The
scattering angle at the target and the momentum of a scattered
particle  were determined by the  ray-tracing method.  The
$^{58}$Ni($\alpha, \alpha^{\prime}$) spectra  were  measured in the
angular range of 0$^\circ$ to 8.5$^\circ$ for two
excitation-energy-bite settings of the spectrometer
($E_{x}$ = 5.0--35.0 MeV and $E_{x}$ = 22.0--52.0 MeV). The
primary beam was stopped at one of four different Faraday cups, depending
on the scattering angle and the excitation energy bite of the
spectrometer.  The vertical position spectrum obtained  in the
double-focused  mode of the spectrometer was exploited to eliminate
the instrumental background due to Coulomb scattering of the beam at
the target and subsequent rescattering by the edges of the entrance
slit, the yoke, and walls of the spectrometer \cite{itoh,uchida1}. Fig.~1 shows
an excitation energy spectrum for the $^{58}$Ni($\alpha$,
$\alpha^{\prime}$) reaction at $\theta_{avg.} = 0.69^{\circ}$  after
subtraction of the instrumental background. A
prominent ``bump'' corresponding to (ISGMR + ISGQR) in $^{58}$Ni is
observed at $\it{E}_{\rm{x}}$ = 10--25 MeV and another bump
[ISGDR + the high-energy octupole resonance (HEOR)] is visible as a shoulder at  $\it{E}_{\rm{x}} \sim$ 33
MeV. There is an underlying continuum in the high excitation-energy
region in the  spectrum. Since there is no sound theoretical basis to
estimate and subtract the physical continuum from the excitation
energy spectrum, it is reasonable to assume that the  continuum
background  remaining after
elimination  of  the instrumental background is the contribution
from the  higher multipoles and the three-body channels resulting, for example,
from knock-out reaction.  In the present work, a
multipole-decomposition (MD) analysis has been performed to extract giant
resonance strengths, by taking
into account the transferred angular momentum up to $\Delta L$ = 7.  The
cross-section data were binned in 1-MeV energy intervals to reduce the
statistical fluctuations. For each excitation-energy bin from 10.5
MeV to 49.5 MeV, the experimental angular distribution
$\sigma^{exp}(\theta_{c.m.}, E_{x})$ has been fitted by means of the
least-square method with the linear combination of calculated
distributions $\sigma^{cal}(\theta_{c.m.},E_{x})$ defined by:

\begin{equation}
\sigma^{exp}(\theta_{c.m.}, E_{x}) = \sum^{L=7}_{L=0}
a_{L}(E_{\rm{x}}) \times \sigma^{cal}_{L}(\theta_{c.m.}, E_{x}),
\end{equation}
where $\sigma^{cal}_{L}(\theta_{c.m.}, E_{x})$ is the calculated
distorted-wave Born approximation (DWBA) cross section corresponding
to 100\% energy-weighted sum rule (EWSR) for the $L^{th}$ multipole.

The DWBA calculations were performed following the method of Satchler
and Khoa \cite{satch1} using density-dependent single folding,
with a Gaussian $\alpha$-nucleon potential (range $t$~=~1.88~fm) for the
real part, and a
Woods-Saxon imaginary term; the calculations were carried out with the computer
code PTOLEMY \cite{ptol}. Input parameters for PTOLEMY were
modified \cite{satch2} to take into account the correct relativistic
kinematics. The shape of the real part of the potential and the form
factors for PTOLEMY were obtained using the codes SDOLFIN and
DOLFIN \cite{dolf}. We used the transition densities and sum rules for
various multipolarities described in Refs. \cite{hara,satch3}. The radial
moments for $^{58}$Ni were obtained by numerical integration of the
Fermi mass distribution with $c$ = 4.08 fm and $a$ = 0.515 fm \cite{satch3}.
The folding-model parameters with the computer code PTOLEMY were obtained
from analysis of  $^{58}$Ni+ $\alpha$  elastic- and $J^{\pi}$ = 2$^+$
inelastic-scattering data at $E_{\alpha}$ = 386 MeV taken in a separate
experiment. The folding model parameter extracted for the real part of the
potential is V = 37.02 MeV, and the parameters for the Woods-Saxon type
imaginary part were:  W = 36.86 MeV, $r_{I}$(reduced radius) = 0.95 fm,
and $a_{I}$(diffuseness) = 0.67 fm. Using these
parameters, the DWBA calculation for the first  $J^{\pi}$ = 2$^+$ state
in $^{58}$Ni was carried out with PTOLEMY using a collective form
factor with the previously-known $B(E2)$ = 0.070 e$^2$b$^2$ \cite{ram,spear}.
Fig.~2 compares the
results of the calculations and the experimental data; the
calculations reproduce elastic scattering cross sections as well as the
inelastic scattering differential cross section for the 2$^{+}$ state very
well.

The contribution  of the IVGDR excitation to the measured cross sections
was subtracted prior to multipole decomposition. The cross section
for IVGDR excitation was calculated using the strength distribution
obtained from  photonuclear work \cite{berm} in conjunction with DWBA
calculations on the basis of the Goldhaber-Teller model. The fits  of
the angular distributions for two energy bins near the peaks of the
ISGMR and ISGDR are shown in Fig.~3; the $L$ = 0, 1, 2 and 3 contributions to
the differential cross section are also
shown. The ISGMR has a maximum at $\theta_{\rm{c.m.}} = 0^{\circ}$
and its contribution is dominant in comparison to the other multipoles at
$E_{x}$ = 18.5 MeV. Similarly, the ISGDR has a
maximum at $\theta_{\rm{c.m.}} \sim 2^{\circ}$ and its contribution is dominant
in comparison to the other multipoles at 29.5 MeV.  The fitted parameters
$a_{L}(E_{x})$ so obtained are fractions of the EWSR's, which can be
related to the strengths $S_{L}(E_{x})$ as follows:
\begin{mathletters}
\begin{equation}
S_{\rm{0}}(E_x) = {2\hbar^2A<r^2> \over mE_x}a_{\rm{0}}(E_x), \\
\label{eq:2a}
\end{equation}
\begin{equation}
S_{1}(E_x) = {3\hbar^2A\over 8\pi mE_x}(11<r^4>-{25 \over 3}<r^2>^2-10\epsilon <r^2>)  a_{1}(E_x),
\label{eq:2b}
\end{equation}
\begin{equation}
S_{\ge 2}(E_x) = {\hbar^2A \over 8\pi mE_x}L(2L+1)^2<r^{2L-2}>a_{\rm{L}}(E_x),
\label{eq:2c}
\end{equation}
\end{mathletters}

\noindent
where m, A, and $<r^N>$ are the
nucleon mass, the mass number, the N$^{th}$ moment of the ground-state density, respectively, and
$\epsilon$=(4/E$_{2}$+5/E$_{0}$)$\hbar^2$/3mA. E$_{\rm{0}}$ and E$_{2}$
are the centroid energies of the GMR and GQR, respectively.
The strength
distributions extracted from these fits for $L$ = 0 (ISGMR), $L$ = 1 (ISGDR),
and $L$ = 2 (ISGQR) in $^{58}$Ni  are shown in Fig.~4. In
order to examine the reliability of the strength distributions
obtained from the fits in the MD analysis, we varied the $L_{max}$
value from $L$ = 6 to $L$ = 8. However, the extracted strength
distributions for $L$= 0--3 did not change in any significant way. In addition,
a completely independent data analysis, using a different folding-model
potential, led to essentially the same results for the various strength
distributions.  

The centroid energy of ISGMR, shown in Fig.~4(a), was determined  to be
$E_{x}$ = 19.9$^{+0.7}_{-0.8}$ MeV between $E_{x}$ = 10.5 and 32.5 MeV.
A total of 92$^{+4}_{-3}\%$ of the $E0$ EWSR was identified
in the above excitation-energy range. [The errors quoted in all EWSR values
here are only statistical; in addition, there may be a 15--20\% systematic
error in the EWSR fractions because of the uncertainties associated with 
the DWBA calculations used in the MDA analysis.] This result is similar to that
reported earlier \cite{lui}, where the fraction of  74$^{+22}_{-12} \%$
for the $E0$ EWSR value is observed between   $E_{x}$ = 12.0 to 31.1 MeV
with a centroid of 20.30$^{+1.69}_{-0.14}$ MeV.

The strength distribution of ISGDR is shown in Fig.~4(b). We observe
a ``bi-modal'' strength distribution between $E_{x}$ = 10.5 and 42.5
MeV. A low-energy (LE) component at  $\sim$16 MeV appears as a
shoulder at the low-energy side of the extracted ISGDR strength. The excitation
energy of this component is much higher than that expected
for the previously-mentioned $1\hbar \omega$ component of the
ISGDR \cite{isgdr1}, and lower than that of the IVGDR ($E_{x}$ = 18 MeV
\cite{berm}). This is the first
observation of concentrated ``bi-modal'' isoscalar $E1$ resonance in an
A $<$ 90 nucleus. In the previous
measurement on ISGDR strength distribution in $^{58}$Ni \cite{lui}, it was
reported that 41\%
$E1$ EWSR was spread more or less uniformly over $E_{x}$ = 12 to 35 MeV. In the
present case, not only is the ISGDR strength concentrated and has
a ``bi-modal''
distribution similar to that observed in the A $\ge$ 90 nuclei, it would
appear that nearly all of the expected ISGDR strength has been observed:
A total of 98$^{+4}_{-5}\%$ of the $E1$ EWSR was
identified between $E_{x}$ = 20.5 and 40.5 MeV. The centroid
energy of the ISGDR for the same excitation energy range is determined to be
30.8$^{+1.7}_{-1.1}$ MeV.
The difference between the present result and the result of Ref. \cite{lui}
might be attributable to the fact that in Ref. \cite{lui} the
multipole decomposition is carried out after
subtracting a ``background'' from the excitation energy spectrum,
whereas, as pointed out earlier, no such subtraction
is required in the present analysis since our spectra have been rendered free
of all instrumental background.

The ISGQR strength distribution is shown in Fig.~4(c). In this case,
73$^{+3}_{-3}\%$ of the  $E2$
EWSR value has been observed in the excitation energy range
between  $E_{x}$ = 10.5--21.5 MeV. The
centroid energy of ISGQR is determined to be 16.3$^{+0.8}_{-0.9}$. This result
is consistent with the result of Ref. \cite{lui}, where $E2$ strength
corresponding to 115$ \pm$18\% of
the $E2$ EWSR was found between $E_{x}$ = 10.5--20.5 MeV with a centroid of
16.1$\pm$0.3 MeV and rms width of 2.4$\pm$0.2 MeV. The
ISGQR strength shows a near constant value beyond $E_{x}$ = 20 MeV. At
present, the reasons behind this extra strength
are not fully understood.
Similarly, enhanced $E1$ strengths at high excitation energies have been noted
previously in other nuclei \cite{uchida2,itoh},
and have been attributed to contributions to the continuum from three-body
channels, such as knock-out reactions. These processes are only
implicitly included in the MD analysis as background and may lead to spurious
contribution to the extracted giant resonance strengths
at higher energies where the associated cross sections are very small. This
conjecture is supported by recent charged-particle decay measurements on ISGDR
wherein no such spurious strength at high excitation energies is
observed \cite{garg,matyas}.
Incidentally, a similar
increase at higher excitation energies has also been reported
recently in $E0$ strength in $^{12}$C \cite{bency}, when a multipole
decomposition was carried out
without subtracting the continuum from the excitation-energy spectra.

To get a quantitative understanding of ISGDR strength distribution,
the experimental strength distribution has been compared with
the predictions of  quasi-particle  random phase approximation
(QRPA) \cite{colo1}
as shown in Fig.~5. The agreement between the experimental and
theoretical ISGDR strength distributions is rather good, 
except at the very highest excitation energies where, as mentioned previously,
the experimentally
extracted ISGDR strength is compromised by the limitations of the MD analysis procedure.
This is quite remarkable since, in general, the details of the theoretical
strength distributions do not quite match the experiment even when the
centroid energies are in agreement with the experimental data. 

In summary, we have performed $^{58}$Ni($\alpha, \alpha^{\prime}$)
measurements at $E_{\alpha}$ = 386 MeV to study the excitation of ISGDR.
The ISGDR strength distribution has been obtained up to $E_{x}$=
49.5 MeV by multipole-decomposition analysis. A two-component  ISGDR
strength distribution has been observed for the first time in
$^{58}$Ni and, indeed, in any $A <$ 90 nucleus.
The centroid energy of the high-energy component of the ISGDR
($E_{x}$ = 30.8$^{+1.7}_{-1.1}$ MeV) is
consistent with the global systematics and the strength
distribution for the HE component of the ISGDR is in good agreement,
qualitatively and quantitatively,
 with predictions of
QRPA calculations with a nuclear matter incompressibility value of
$K_{nm}$ = 217 MeV \cite{colo2}.

This work has been supported in part by the US-Japan Cooperative
Science Program of JSPS and the US National Science Foundation (grants number
 INT03-42942 and PHY04-57120).

\newpage

\begin{figure}
\includegraphics[width=10.5cm]{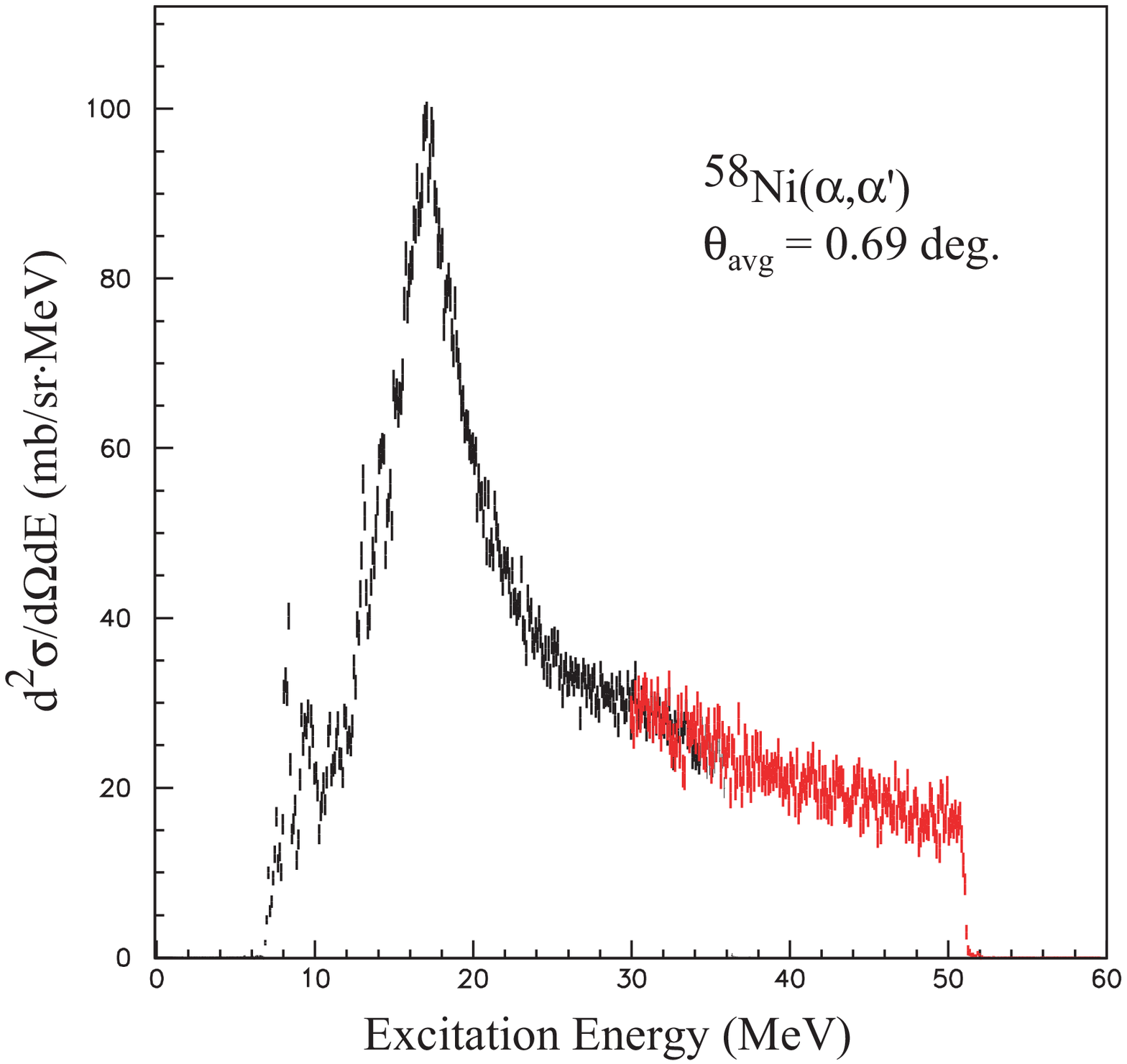}
\caption{Excitation energy spectrum for the
$^{58}$Ni($\alpha,\alpha^{\prime}$) reaction at $E_{\alpha}$= 386
MeV. Inelastically-scattered $\alpha$ particles were measured with the
magnetic spectrometer at $\theta = 0^{\circ}$ with  
two different settings of the magnetic field
to cover the excitation-energy ranges of $E_{x}$=5.0--35.0 MeV
and $E_{x}$= 22.0--52.0 MeV.}
\label{fig1}
\vspace{5cm}
\end{figure}

\newpage
\begin{figure}
\includegraphics[width=8.5cm]{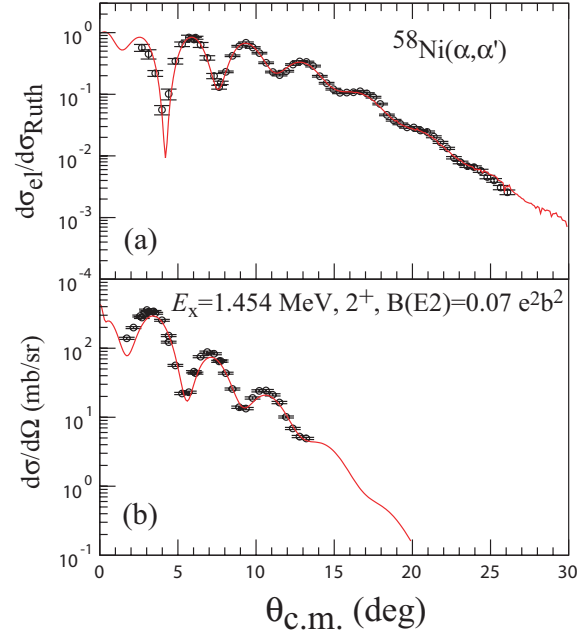}
\caption{Cross sections for  (a) the $\alpha$ + $^{58}$Ni  elastic
scattering and (b) the $^{58}$Ni$(\alpha,
\alpha^{\prime})$$^{58}$Ni(2$^+$) reaction at $E_{\alpha}$ = 386 MeV.
The solid lines are the results of the folding model calculations.}
\label{fig2}
\end{figure}

\newpage
\begin{figure}
\includegraphics[angle=0, width=8.5cm]{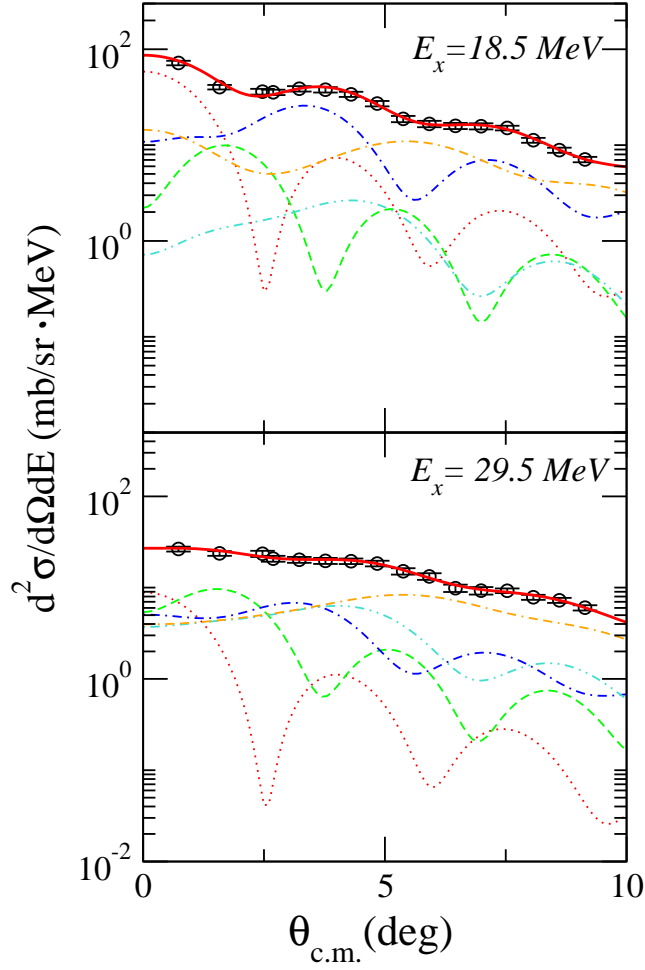}
\caption{Angular distributions of selected 1 MeV bins for the  $
^{58}$Ni($\alpha, \alpha^{\prime}$) reaction at 386 MeV. (a) Results
for $E_{x}$ = 18.5 MeV. The open circles are the experimental
data. The lines show contributions from $L = 0$ (dotted line), $L = 1$
(dashed line), $L = 2$ (dot-dashed line), $L = 3$ (double dot-dashed
line) and other higher-multipole components including IVGDR (double
dash-dotted line), respectively. (b) Same as part (a), except for
$E_{x}$ = 29.5 MeV.}
\label{fig3}
\end{figure}

\newpage
\begin{figure}
\includegraphics[angle=0, width=10.5cm]{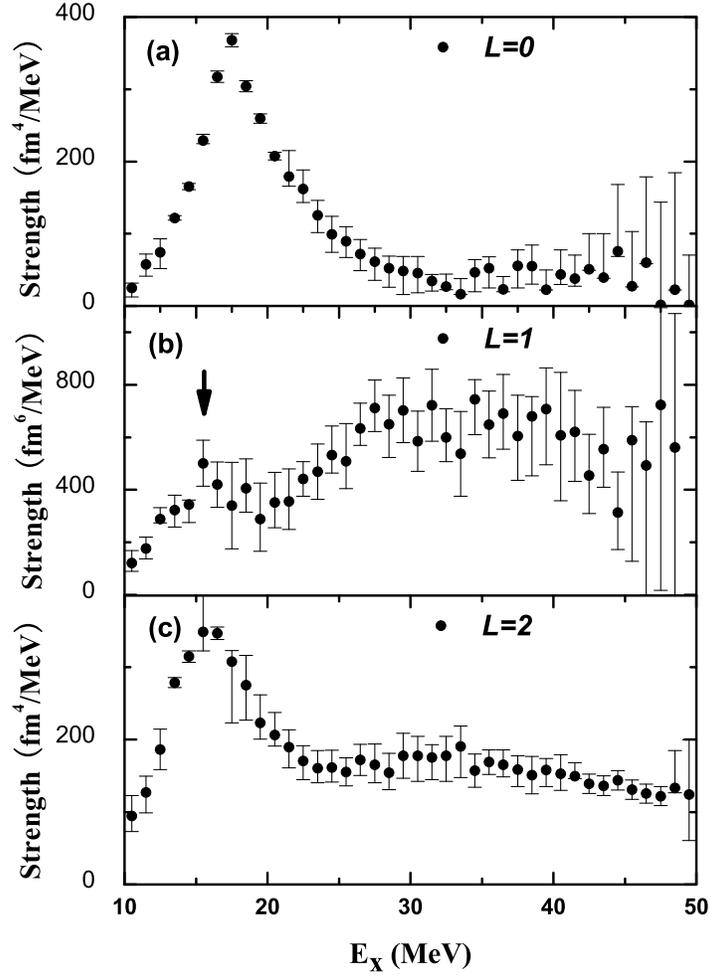}
\caption{Strength distributions for the $L$ = 0, 1, and 2
in $^{58}$Ni. (a) ISGMR, (b) ISGDR, (c) ISGQR.
The errors shown for each excitation-energy bin were estimated by
changing the strength parameter for one component in order to satisfy
$\Delta\chi^{2}$ increase by 20\% while fitting with the other parameters 
remaining free. The low-energy component of the ISGDR is indicated by an arrow.
}
\label{fig4}
\end{figure}
\newpage
\begin{figure}
\includegraphics[angle=0, width=12.5cm]{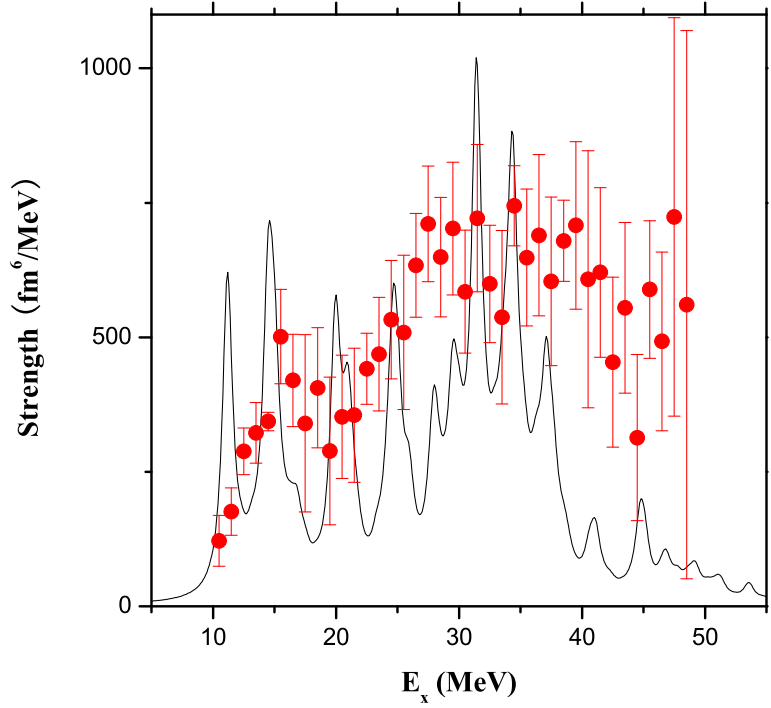}
\caption{Comparison of the  experimental ISGDR strength distribution in
$^{58}$Ni with the predictions of recent QRPA calculation (continuous line)
\cite{colo1}.}
\label{fig5}
\end{figure}
\end{document}